\documentclass[reprint,aps,pra,superscriptaddress]{revtex4-1}
\usepackage{amssymb,amsmath}
\usepackage{graphicx}
\usepackage{dcolumn}
\usepackage{multirow}
\usepackage{color}
\usepackage[english]{babel}
\usepackage[caption=false]{subfig} 

\newcommand{\ket}[1]{\ensuremath{|{#1}\rangle}}

\begin{document}

\def\simlt{\mathrel{\lower .3ex \rlap{$\sim$}\raise .5ex \hbox{$<$}}}

\title{\textbf{\fontfamily{phv}\selectfont 
 High fidelity resonant gating of a silicon based quantum dot hybrid qubit}}
\author{Dohun Kim}
\author{D. R. Ward}
\author{C. B. Simmons}
\affiliation{Department of Physics, University of Wisconsin-Madison, Madison, WI 53706}
\author{D. E. Savage}
\author{M. G. Lagally}
\affiliation{Department of Materials Science and Engineering, University of Wisconsin-Madison, Madison, WI 53706, USA}
\author{Mark Friesen}
\author{S. N. Coppersmith}
\author{M. A. Eriksson}
\affiliation{Department of Physics, University of Wisconsin-Madison, Madison, WI 53706}

\begin{abstract}
We implement resonant single qubit operations on a semiconductor hybrid qubit hosted in a three-electron Si/SiGe double quantum dot structure. By resonantly modulating the double dot energy detuning and employing electron tunneling-based readout, we achieve fast ($>$ 100 MHz) Rabi oscillations and purely electrical manipulations of the three-electron spin states. We demonstrate universal single qubit gates using a Ramsey pulse sequence as well as microwave phase control, the latter of which shows control of an arbitrary rotation axis on the X-Y plane of the Bloch sphere. Quantum process tomography yields $\pi$ rotation gate fidelities higher than 93 (96) percent around the X(Z) axis of the Bloch sphere. We further show that the implementation of dynamic decoupling sequences on the hybrid qubit enables coherence times longer than 150 ns.  
\end{abstract}

\maketitle
Isolated spins in semiconductors provide a promising platform to explore quantum mechanical coherence and develop engineered quantum systems~\cite{Hanson:2007p1217,Zwanenburg:2013p961,Loss:1998p120,Kane:1998p133,Elzerman:2004p431,Petta:2005p2180,Koppens:2006p766,Foletti:2009p903,Laird:2010p1985,Gaudreau:2011p54,Pla:2012p489,Medford:2013p654,Buch:2013p2017}.
Silicon has attracted great interest as a host material for developing spin qubits because of its weak spin-orbit coupling and hyperfine interaction, and several architectures based on gate defined quantum dots have been proposed and demonstrated experimentally~\cite{Maune:2012p344,Kawakami2014}. Recently, a quantum dot hybrid qubit formed by three electrons in double quantum dot was proposed~\cite{Shi:2012p140503,Koh:2012p250503}, and non-adiabatic pulsed-gate operation was implemented experimentally~\cite{Kim:2014nature}, demonstrating simple and fast electrical manipulations of spin states with a promising ratio of coherence time to manipulation time. However, the overall gate fidelity of the pulse-gated hybrid qubit is limited by relatively fast dephasing due to charge noise during one of the two required gate operations. Here we perform the first microwave-driven gate operations of a quantum dot hybrid qubit, avoiding entirely the regime in which it is most sensitive to charge noise. Resonant detuning modulation along with phase control of the microwaves enables a $\pi$ rotation time of less than 5 ns (50 ps) around X(Z)-axis  with high fidelities $>$ 93 (96) \%. We also implement Hahn echo \cite{Koppens:2008p236802, Vanderypen:2005p1037, Dial:2013p146804} and Carr-Purcell (CP) \cite{Bluhm:2011p109} dynamic decoupling sequences with which we demonstrate a coherence time of over 150 ns. We further discuss a pathway to improve gate fidelity to above 99\%, exceeding the threshold for surface code based quantum error correction~\cite{Fowler:2012p032324}.

\begin{figure*}
\includegraphics[width=1\textwidth]{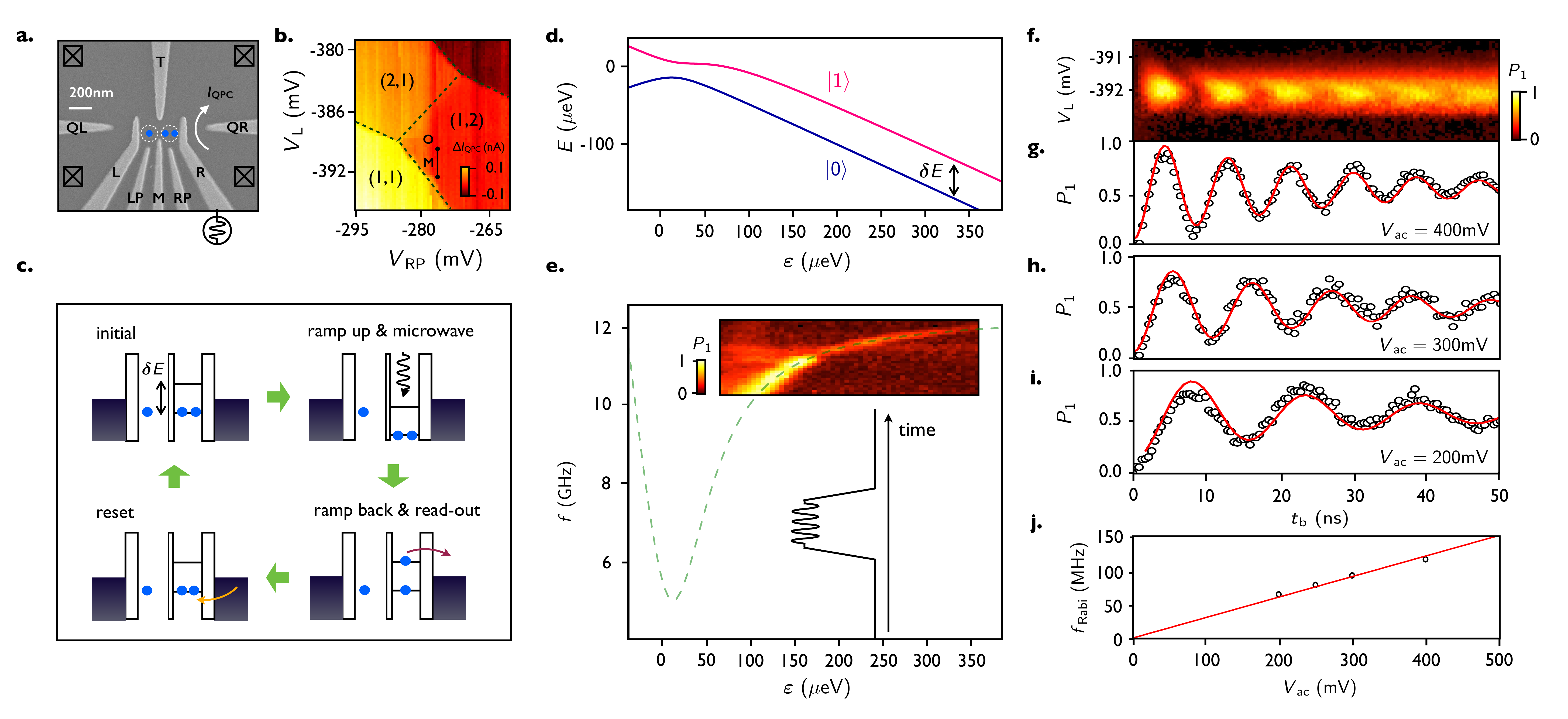}
\caption{
\textbf{Microwave-driven coherent manipulation and readout of a hybrid qubit in a Si/SiGe double quantum dot device. a}, 
SEM image and schematic labeling of a device lithographically identical to the one used in the experiment.  \textbf{b}, Charge stability diagram near the (1,1)-(2,1)-(1,2) charge transition, showing the gate voltages used for microwave manipulation (O) and measurement (M). For clarity, linear background slope was removed from the raw charge sensing data.
\textbf{c}, Schematic description of the qubit initialization, manipulation, readout, and reset processes. 
\textbf{d}, Energy E as a function of detuning $\varepsilon$ for the qubit states, calculated with Hamiltonian parameters measured in Ref. \cite{Kim:2014nature}. \textbf{e}, (inset) Probability $P_1$ of the state to be $\ket{1}$ at the end of the driving sequence shown as a function of $\varepsilon$ and the excitation frequency $f$ of the microwave applied to gate R. In the main panel, the dashed green curve is the energy difference between the ground state and the lowest-energy excited state, as determined in~\cite{Kim:2014nature}. 
\textbf{f-j}, Coherent Rabi oscillation measurements. \textbf{f}, $P_1$ as a function of the voltage $V_\text{L}$ and the microwave pulse duration $t_\text{b}$ with $f$=11.52 GHz and excitation amplitude $V_\text{ac}=400$~mV. 
\textbf{g}, Linecut of $P_1$ near $V_\text{L}=$-392 mV, showing $\approx$ 110 MHz Rabi oscillations. The red solid curve is a fit to an exponentially damped sine wave with best fit parameter $T_{2}^*$ = 33 ns. \textbf{h-i}, Rabi oscillation data with microwave amplitude 300mV (\textbf{h}) and 200mV (\textbf{i}).
\textbf{j}, Rabi oscillation frequency $f_\text{Rabi}$ as a function of $V_\text{ac}$ with fixed $f$ =11.52 GHz. The good agreement of a linear fit (red line) to the data is strong evidence that the measured oscillations are indeed Rabi oscillations, with the Rabi frequency proportional to the driving amplitude.}

\label{fig1} 
\end{figure*}

The quantum dot hybrid qubit combines desirable features of charge (fast manipulation) and spin (long coherence time) qubits.  The qubit states can be written as $\ket{0}=\ket{\downarrow}|S\rangle$, where $S$ denotes a singlet state in the right dot, and
$\ket{1}=1/\sqrt{3}\ket{\downarrow}|T_0\rangle-\sqrt{2/3}\ket{\uparrow}|T_-\rangle$, where $T_0$ and $T_-$ are two of the triplet states in the right dot. The states $\ket{0}$ and $\ket{1}$ have the nearly same dependence on $\varepsilon$ in the range of detuning at which the qubit is operated (see Figs.~1c,d), enabling quantum control that is largely insensitive to charge fluctuations. Moreover, electric fields couple to the qubit states
and enable high speed manipulation~\cite{Shi:2012p140503,Koh:2012p250503,Ferraro:2014p1,Mehl:2015p035430,Ferraro:2015p47,deMichielis:2015p065304} . Previously, we experimentally demonstrated non-adiabatic quantum control (DC-pulsed gating) of the hybrid qubit, where the manipulation and measurement scheme required the use of a detuning regime that is sensitive to charge noise (with $\varepsilon$ near but not equal to zero---see Fig. 1d)~\cite{Kim:2014nature}. Moreover, DC-gating requires abrupt changes in detuning. With a given bandwidth in the transmission line, pulse imperfections arising, e.g., from frequency dependent attenuation, lead to inaccurate control of rotation axes. In this work, we demonstrate resonant microwave-driven control and state-dependent tunneling readout of the qubit, which together overcome this limitation of DC-pulsed gating and enable full manipulation on the Bloch sphere at a single operating point in detuning that is well-protected from charge noise.

The experiments here are performed in a double dot with a gate design as shown in Fig.\ 1a and with electron occupations as shown on the stability diagram of Fig.\ 1b. The electron occupations and energy level alignments used for qubit initialization, readout, and microwave spectroscopy of the qubit states are shown schematically in Fig.\ 1c.  All the experiments reported here start with an initial dot occupation of (1,2) and with system in state $\ket{0}$, prepared at a detuning $\varepsilon\approx 230 \mu\text{eV}$; this detuning is also used for measurement and corresponds to point M in Fig.~1b.  After initialization, we apply a microwave burst pattern at point O, which either coincides with point M or is reached through an adiabatic ramp in detuning (the latter case is illustrated in Fig.~1b).
Because of the presence of the third electron in the pair of dots, the hybrid qubit states couples to time varying electric fields~\cite{Shi:2012p140503}, and qubit excitation or rotation occurs when the applied microwave frequency is resonant with the qubit energy level difference.  The measurement point M is chosen so that the Fermi level of the right resorvior is in between the energies of $\ket{0}$ and $\ket{1}$, and we use the qubit state-dependent tunneling to project states $\ket{0}$ and $\ket{1}$ to the (1,2) and (1,1) charge states, respectively.  Waiting at point M for $\approx 10 \mu\text{s}$ also resets the qubit to state $\ket{0}$, by tunneling an electron from the reservoir, if needed. Thus, the qubit state population following the microwave burst is measured by monitoring the current $I_\text{QPC}$ through the charge-sensing quantum point contact (Fig.~1a). Details of the measurement procedure and probability normalization are in Supplementary Information S1.

We perform microwave spectroscopy of the qubit intrinsic frequency---the energy difference $\delta E$ in Fig.~1d---by applying the voltage pulse shown in the inset to Fig.~1e.  The color plot in that figure shows the resulting probability of measuring state $\ket{1}$ after applying this pulse to initial state $\ket{0}$.  The measured resonance and qubit energy dispersion agrees well with the green dashed curve, which is the calculated energy level diagram with Hamiltonian parameters measured in our previous study~\cite{Kim:2014nature}. As is clear from the color plot in Fig.~1e, the linewidth of the resonant peak narrows significantly at $\varepsilon > 200 \mu\text{eV}$, becoming much narrower than the resonance in the charge qubit regime ($\varepsilon\approx 0$)~\cite{Kim2014preprint}.  This linewidth narrowing corresponds to an increase in the inhomogeneous dephasing time, and it is this range in detuning that corresponds to the hybrid qubit regime.  

Applying microwave bursts to gate R in the hybrid qubit regime yields Rabi oscillations, as shown in Figs.~1f-i. The Rabi frequency increases as a function of increasing microwave amplitude $V_\text{ac}$ (measured at the arbitrary waveform generator), resulting in Rabi frequencies as high as 100 MHz. Fig.\ 1j shows the power dependence of the qubit oscillations, revealing an oscillation frequency that is linear in the applied amplitude, as expected for Rabi oscillations. The speed of the X-axis rotation demonstrated here is comparable to electrically manipulated spin rotations in InSb and InAs, which rely on strong spin-orbit coupling of the host material~\cite{Petersson:2012p380,Berg:2013p066806}; here we achieve fast rotations solely through electric field coupling to the qubit states. This coupling is also highly tunable, since it is determined by the ground and excited state inter-dot tunnel couplings~\cite{Koh:2013p19695}. Below, we characterize gates with the qubit frequency chosen to be $\approx$~11.52~GHz.

\begin{figure*}
\includegraphics[width=1\textwidth]{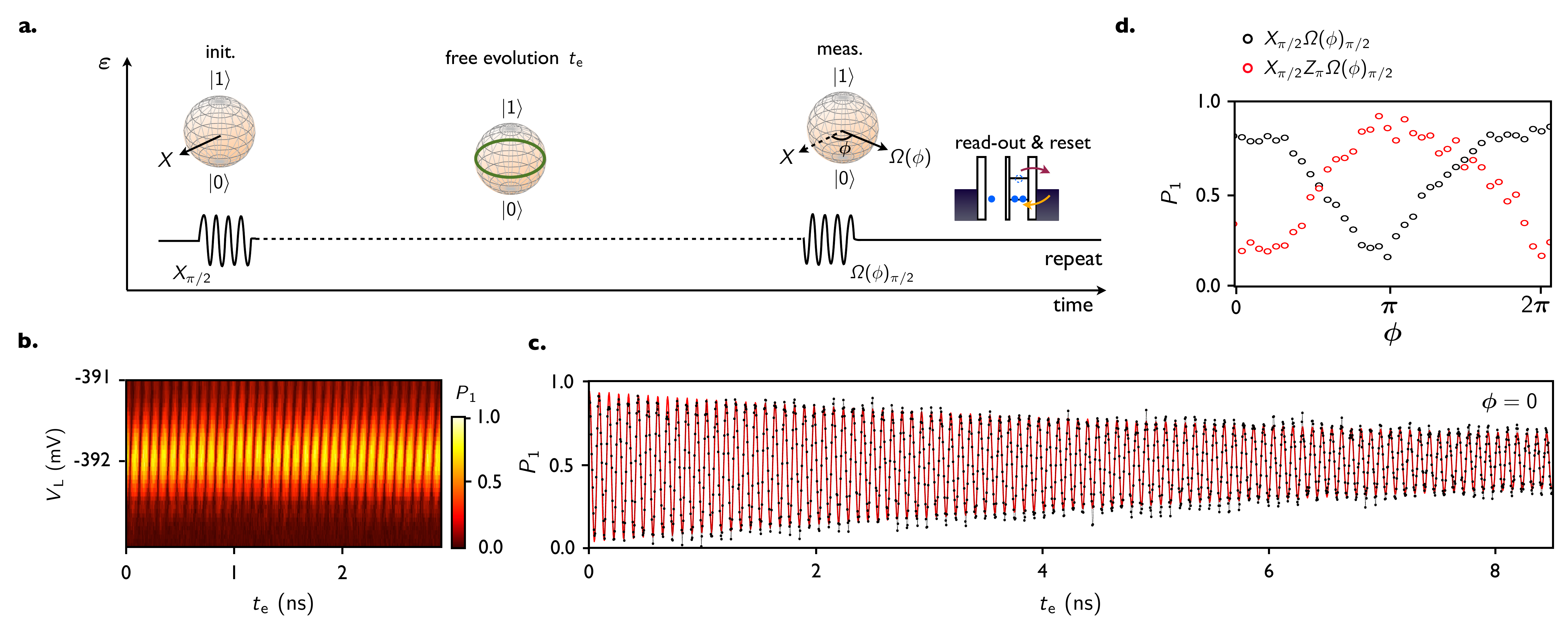}
\caption{
\textbf{Ramsey fringes and two-axis control of the qubit. a}, Schematic diagram of the pulse sequences used to perform universal control of the qubit. Both the delay $t_\text{e}$ and the phase $\phi$ of the second microwave pulse are varied in the experiment.
\textbf{b-c}, Experimental measurement of Z-axis rotation. In a Ramsey fringe (Z-axis rotation) measurement, the first $X_\text{$\pi$/2}$ gate rotates the Bloch vector onto the $X$-$Y$ plane, and the second $X_\text{$\pi$/2}$ gate ($\phi$=0) is delayed with respect to the first gate by $t_\text{e}$, during which time the state evolves freely around the Z-axis of the Bloch sphere. \textbf{b}, $P_\text{1}$ as a function of $V_\text{L}$ and $t_\text{e}$ for states initialized near $|Y\rangle$. \textbf{c}, $P_\text{1}$ as a function of $t_\text{e}$ with fixed $V_\text{L}=-391.7$~mV, showing $\approx 11.5$~GHz Ramsey fringes.  The red solid curve is a damped sine wave with best fit parameter $T_{2}^* =11$~ns. 
\textbf{d}, Effect of the phase $\phi$ of the second microwave pulse on the state $\ket{Y}$ (by applying $X_{\pi/2}$ on $\ket{0}$, black), and $\ket{-Y}$ (by applying $X_{\pi/2}$ and $Z_{\pi}$ on $\ket{0}$, red). The clear oscillation of $P_{1}$ as a function of $\phi$ in both cases demonstrates control over the second rotation axis by control of the phase $\phi$.}
\label{fig2} 
\end{figure*}

We characterize the inhomogeneous dephasing time by performing a Ramsey fringe experiment, which also demonstrates Z-axis rotations on the qubit Bloch sphere. The microwave pulse sequence is shown schematically in Fig.~2a. We first prepare the state $\ket{Y}=\sqrt{1/2}(\ket{0}+i\ket{1})$ by performing an $X_{\pi/2}$ rotation. Z-axis rotation results from the evolution of a relative phase between states $\ket{0}$ and $\ket{1}$, given by $\varphi={-t_\text{e}\delta E /\hbar}$, where $t_\text{e}$ is the time spent between the state preparation and measurement $\text{X}_{\pi/2}$ pulses, the latter of which is used to project the Y-axis component onto the Z-axis. The final probability $P_{1}$ is measured as described above. Fig.~2b shows the resulting quantum oscillations as a function of $V_\text{L}$, which controls the detuning energy, and $t_\text{e}$. Fig.\ 2c shows a linecut taken near the optimal resonant condition ($V_\text{L}\approx - 392$~mV), showing clear oscillations in $P_{1}$ consistent with the qubit frequency of $\approx 11.5$ GHz. By fitting the oscillations to an exponentially damped sine wave (red solid curve), we extract an inhomogeneous dephasing time $T_{2}^*= 11$~ns, consistent with the value measured previously with non-adiabatic pulsed gating on the same device with similar intrinsic qubit frequency \cite{Kim:2014nature}. We estimate the typical tunneling-out time $T_\text{o} \approx 200$~ns (see Supplementary Information S1), so that the inhomogeneous coherence time is not likely limited by electron tunneling to the reservoir during the measurement phase.

Resonant microwave drive also enables arbitrary two-axis control on the $X$-$Y$ plane of the Bloch sphere by varying the relative phase $\phi$ of the $\text{X}_{\pi/2}$ pulses. Fig.~2d shows a measurement of $P_{1}$, demonstrating both two-axis control and phase control. Starting from a maximum (minimum) $P_{1}$ at $\phi$ = 0, when we apply  $X_{\pi/2}$$\Omega_{\pi/2}$ ($X_{\pi/2}$$Z_\pi$$\Omega_{\pi/2}$) on the state $\ket{0}$, $P_{1}$ oscillates as a function of the relative phase $\phi$ that determines the axis of the $\Omega_{\pi/2,\phi}$ rotation. The deviation from an ideal sinusoidal oscillation stems from limited phase resolution of our method of waveform generation (see Supplementary Information S1). 

\begin{figure*}
\includegraphics[width=1\textwidth]{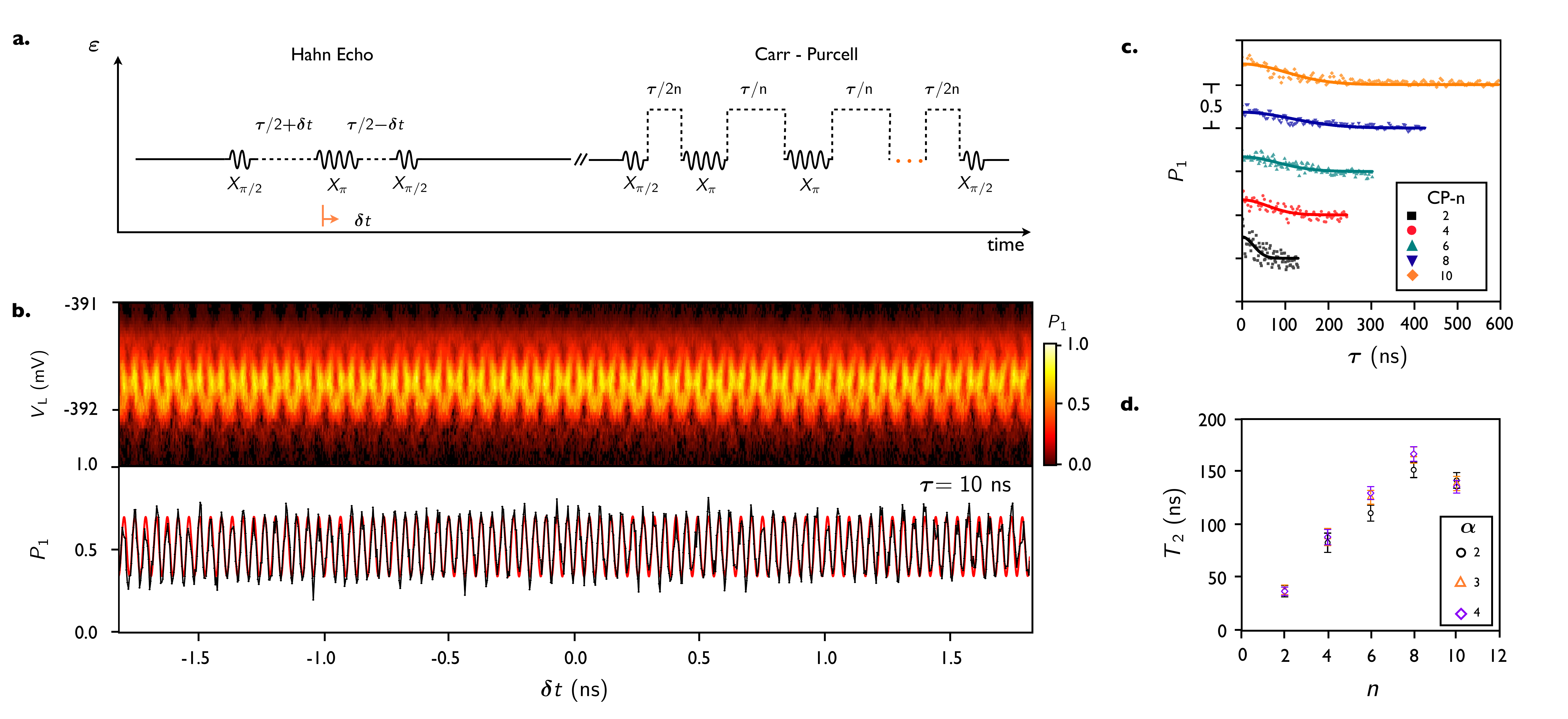}
\caption{
\textbf{Increasing coherence time with dynamic decoupling sequences.} \textbf{a}, Schematic pulse sequence for the measurement of Hahn spin echo and Carr-Purcell (CP) dynamic decoupling sequences that correct for noise that is static on the time scale of the pulse sequence~\cite{Koppens:2008p236802, Vanderypen:2005p1037, Dial:2013p146804}. Note that for CP dynamic decoupling, free evolution is performed at $\approx 60\ \mu eV$ more positive detuning than the readout point, in order to prevent tunneling of the state $\ket{1}$ to the reservoir during the manipulation pulses and free evolution.  
\textbf{b}, Typical Hahn echo measurement with fixed total evolution time $\tau = 10\ $ns, showing $P_{1}$ as a function of $V_{L}$ and delay time $\delta t$ of the X$_{\pi}$ pulse. The bottom panel shows line cut of $P_{1}$ near  $V_{\text{L}}$ = -391.7\ mV. The red solid curve shows a fit to a gaussian decay with fixed inhomogeneous coherence time $T_2^*\approx$11\ ns, obtained from the Ramsey fringe measurement of Fig.\ 2.  The oscillations of $P_1$ as a function of $\delta t$ at twice the Ramsey frequency ($\approx$ 23.04 GHz) is clear evidence of spin echo, where we measure an echo amplitude of $\approx 0.3$ for $\tau = 10$\ ns near $\delta t=0$. \textbf{c}, $P_1$ as a function of $\tau$ with fixed $\delta t=0$. The symbols shows the data, while the solid curves are the fit to the decay form $P_{1}(\tau)$ = 0.5 + A $e^{- (\tau/T2)^{\alpha}}$ with fixed exponent $\alpha=$\ 2 for even numbers {\it n} of decoupling X$_{\pi}$ pulses ranging from 2 to 10.
\textbf{d}, Coherence time $T_2$ as a function of $n$ obtained from the fit of CP decay data to the decay form with fixed exponent $\alpha=$\ 2 (black circles), 3 (orange triangles), and 4 (purple diamonds). The resulting $T_2$ is insensitive to the choice of $\alpha$ to within the uncertainty. Applying the CP dynamic decoupling sequences increases the coherence time by more than an order of magnitude compared to $T_{2}^*\approx11$~ns.
}
\label{fig3} 
\end{figure*}

We now turn to echo and dynamic decoupling pulse sequences. Fig.~3a shows Hahn echo~\cite{Koppens:2008p236802, Vanderypen:2005p1037, Dial:2013p146804} and Carr-Purcell dynamic decoupling (CP)\ \cite{Bluhm:2011p109} pulse sequences. Provided that the source of dephasing fluctuates slowly on the timescale of the electron spin dynamics, inserting an $X_{\pi}$ pulse between state initialization and measurement, which is performed with $X_{\pi/2}$ gates, corrects for noise that arises during the time evolution. Fig.~3b shows a typical echo measurement. While keeping the total free evolution time $\tau$ fixed at 10 ns, we sweep the position of the decoupling $X_{\pi}$ pulse to reveal an echo envelope~\cite{Dial:2013p146804, Shi:2013p075416}. In Fig.~3b, the oscillations of $P_{1}$ as a function of $\delta t$ are at twice the Ramsey frequency ($2f_\text{Ramsey}\approx$ 23 GHz) and are well-fit by a gaussian decay (red solid curve). With $\tau$ = 10 ns, we observe a maximum echo amplitude $\approx$ 0.3 near $\delta t \approx$ 0. The gaussian decay indicates that a significant component of the dephasing arises from low-frequency noise processes.

Improvement in coherence times can be obtained by implementing CP sequences (see Fig. 3a), which use multiple $X_{\pi}$ pulses inserted during the free evolution. Since the time scale for the CP sequence is typically longer than $T_\text{o} \approx 200$~ns, an adiabatic detuning offset of amplitude $\approx$ 60 $\mu$eV was applied during free evolution in order to prevent electron tunneling to the reservoir. In the absence of dephasing, the CP sequence with an even number of $X_{\pi}$ pulses applied on the state $\ket{0}$ yields $P_{1}$ = 1. The measured $P_{1}$ as a function of $\tau$, shown in Fig. 3c, decays exponentially due to dephasing: $P_{1}(\tau)$ = 0.5 + A $e^{- (\tau/T_2)^{\alpha}} $, where $\alpha$ depends on the frequency spectrum of the dominant noise sources \cite{Barthel:2010p266808}.  Fig.\ 3d shows the results of fits as a function of the number {\it n} of decoupling $\text{X}_{\pi}$ pulses with
fixed $\alpha$ = 2, 3, and 4.  The resulting coherence time $T_{2}$ shows more than an order of magnitude improvement ($>$ 150 ns) with {\it n} = 8, and the resulting times are approximately independent of the $\alpha$ used in the fit. Beyond {\it n} = 8 we typically observe a decrease in $T_{2}$, which is likely due to accumulation of pulse imperfections in the microwave or detuning pulses. We expect that optimization of microwave pulses can increase $T_{2}$ further.

\begin{figure*}
\includegraphics[width=1\textwidth]{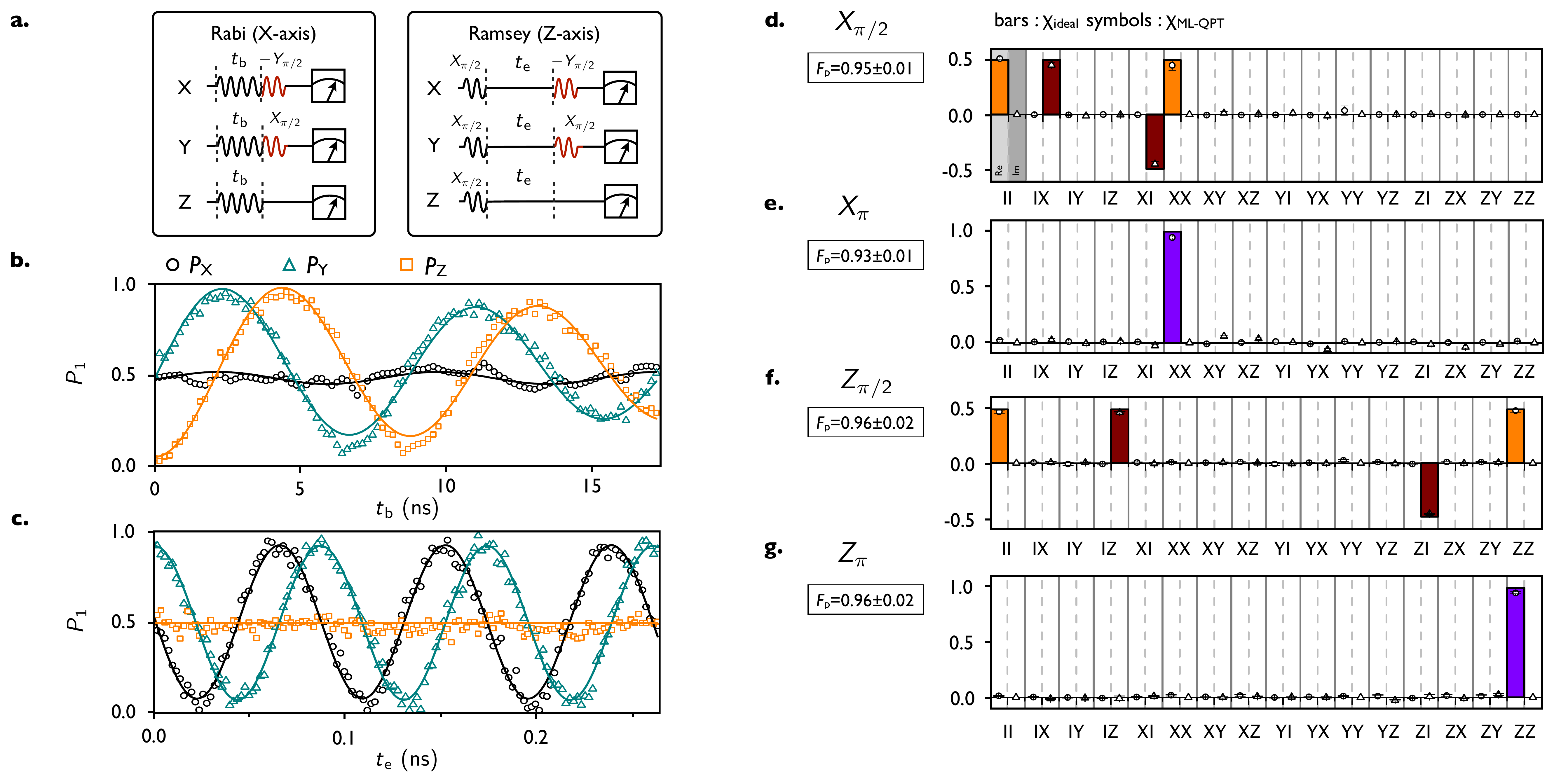}
\caption{
\textbf{State and quantum process tomography (QPT) of the ac-driven hybrid qubit. a}, Schematic of microwave pulse sequences used for the tomographic characterization of continuous Bloch vector evolution under X (left panel) and Z (right panel) axis rotations. \textbf{b-c}, X-axis projection $P_\text{X}$ (green), Y-axis projection $P_\text{Y}$ (orange), and Z-axis projection $P_\text{Z}$ (black) of the Bloch vector evolution under X (\textbf{b}) and Z (\textbf{c}) axis rotations performed in rotating (lab) frame (see main text for more discussion). 
\textbf{d-g}, Real and imaginary parts of the process matrices $\chi$~\cite{Nielsen:2000} in the Pauli basis $\{$\text{I}$, X, Y, Z\}$ for processes $X_{\pi/2}$ ($\textbf{d}$), $X_{\pi}$ ($\textbf{e}$),  $Z_{\pi/2}$ ($\textbf{f}$), and $Z_{\pi}$ ($\textbf{g}$) obtained by quantum process tomography with maximum-likelihood estimation~\cite{Nielsen:2000,Chow:2009p090502, Kim:2014nature, Chow:2009p090502} (real: open circles, imaginary: open triangles) compared to corresponding ideal processes (bars). The error bars represent the standard deviation of the result obtained by 10 distinct input and output density matrices chosen from the state tomography data. Process fidelities defined by $F_\text{p}=Tr(\chi_\text{ideal} \chi_\text{ML-QPT})$ are 93\% for $X_{\pi}$ and 96\% for $Z_{\pi}$ processes, respectively (left panels of \textbf{e} and \textbf{g}).  }
\label{fig4} 
\end{figure*}

 We now present tomographic characterization of the microwave-driven hybrid qubit. To reconstruct the time evolution of the single qubit density matrix, we use the microwave pulse sequences shown schematically in Fig.\ 4a to perform repeated state evolution under an X(Z) gate and perform independent X, Y, and Z axis projective measurements. For state tomography under X(Z)-axis rotation, we prepare initial states near $\ket{0}$ and near $\ket{Y}$. After time evolution under the gate operation, we measure X, Y, and Z axis projections of the time-evolved Bloch vector using $-Y(\pi/2)$, $X(\pi/2)$, and identity operations, respectively, and measure the resulting $P_{1}$. Note that the pulse sequences shown in Fig.\ 4a represent state tomography in the rotating (laboratory) frame for Rabi (Ramsey) oscillations, because the phase of the second $\pi/2$ pulses for the Rabi oscillation tomography evolves as the length of Rabi manipulation $t_\text{b}$ is increased, while the second microwave pulse in the state tomography of the Ramsey fringes has fixed relative phase with respect to the first microwave pulse. For comparison, we also present the state tomography of the Rabi oscillation performed in the lab frame in Supplementary Information S2. Figs.~4b and 4c show X (black circles), Y (green triangles), and Z (orange squares) axis projections of the time-evolved Bloch vector under continuous X (b) and Z (c) axis rotation gates.

Based on the density matrices obtained from the state tomography, we implement quantum process tomography (QPT) to  extract fidelities of single qubit gates on the ac-driven hybrid qubit through the relation~\cite{Nielsen:2000,Chow:2009p090502, Kim:2014nature, Chow:2009p090502},
\begin{equation} 
{\cal E}(\rho)=\sum_{m,n=1}^{4}\tilde{E}_m\rho\tilde{E}^\dagger_n\chi_{mn} ,
\end{equation}
\noindent
where ${\cal E}(\rho)$ is the density matrix specifying the output for a given input density matrix $\rho$, the $\tilde{E}_m$ are the basis operators in the space of $2\times 2$ matrices, and $\chi$ is the process matrix. Experimentally, four linearly independent input and output states are chosen from continuous evolution of the state under X and Z axis rotation available from the state tomography data set, and the maximum likelihood method ~\cite{Chow:2009p090502, Kim:2014nature, Chow:2009p090502} is used to determine $\chi$. Figs.~4d-g show the results of QPT (symbols) performed on the $\pi/2$ and $\pi$ rotations around the X and Z axes and comparison to corresponding ideal rotation process matrices (bars). The error bars of length $\approx$ 0.01 to 0.02 represent the standard deviation of the experimental result obtained by 10 distinct input and output density matrices chosen from the state tomography data. The process matrices $\chi$ obtained from QPT in the Pauli basis $\{\text{I},\sigma_\text{x},\sigma_\text{y},\sigma_\text{z}\}$ yield process fidelities $F_\text{p}=Tr(\chi_\text{ideal} \chi)$ of 93\% and 96\% for $\pi$ rotations around X and Z axes, respectively. Comparing these results to the process fidelities of 85\% and 94\% for X and Z-axis rotation reported previously for the non-adiabatic DC-pulse gated hybrid qubit~\cite{Kim:2014nature}, we find more than a factor of two reduction in the X-axis rotation infidelity.
 
The improvement in overall fidelity of the AC-gated quantum dot hybrid qubit demonstrated here compared to DC-pulsed gating stems mainly from (1) elimination of the need to enter the regime in which the qubit is sensitive to charge noise by using resonant manipulation and tunneling-based readout, and (2) reduced rotation axis and angle errors because resonant driving with fixed frequency enables more accurate control of these quantities. The AC driving in this work was performed by resonantly modulating the energy detuning between the dots. For this type of modulation, the ratio of manipulation time (Rabi period) to coherence time depends strongly on the strength of ground and excited state tunnel couplings~\cite{Koh:2013p19695}. Thus we expect that further fidelity improvement can be achieved by adjusting tunnel couplings. Moreover, recent theoretical work suggests that dynamically modulating tunnel coupling instead of detuning can enable Rabi frequencies exceeding 1~GHz while keeping long coherence times, enabling achievement of gate fidelites exceeding 99\%~\cite{Wong:2014unpublished}. Finally, implementation of dynamic decoupling sequences can be useful to understand the noise spectrum of the system as well as to further enhance the coherence time.

\noindent \textbf{Methods}

The details of the Si/SiGe double quantum dot device are presented in Refs.~\cite{Shi:2013p075416,Simmons:2011p156804}. We work in the region of the charge stability diagram where the valence electron occupation of the double dot is $(1,1)$ or $(1,2)$, as confirmed by magnetospectroscopy measurements~\cite{Simmons:2011p156804,Shi:2014p3020}. All manipulation sequences, including the microwave bursts, are generated by a Tektronix 70002A arbitrary waveform generator and are added to the dot-defining dc voltage through a bias tee (Picosecond Pulselabs~5546-107) before being  applied to gate R. We map the state $\ket{0}$ and $\ket{1}$ to (1,2) and (1,1) charge occupation states respectively, leading to conductance changes through the quantum point contact (QPC).  We measure with a lock-in amplifier (EG\&G model~7265) the difference in conductance with and without the applied microwave burst. When converting time averaged conductance differences to the reported probabilities, tunneling between the (1,2) and (1,1) charge states during the measurement phase is taken into account using the measured times for tunneling out of ($T_\text{o}\simeq $~200 ns) and into ($T_\text{i}\simeq $~2.1 $\mu$s) the dot. Supplementary Information S1 presents the details of the measurement technique and the probability normalization.

\emph{Acknowledgements}
This work was supported in part by ARO (W911NF-12-0607) and NSF (PHY-1104660). Development and maintenance of the growth facilities used for fabricating samples is supported by DOE (DE-FG02-03ER46028). This research utilized NSF-supported shared facilities at the University of Wisconsin-Madison.

\emph{Author Contributions}
DK performed electrical measurements, state and process tomography, and analyzed the data with MAE, MF and SNC. DRW developed hardware and software for the measurements. CBS fabricated the quantum dot device. DES and MGL prepared the Si/SiGe heterostructure. All authors contributed to the preparation of the manuscript. 

\emph{Additional Information}
Supplementary information accompanies this paper. Correspondence and requests for materials should be addressed to Mark A. Eriksson (maeriksson\emph{@}wisc.edu).

\renewcommand{\theequation}{S\arabic{equation}}
\setcounter{equation}{0}
\renewcommand{\thefigure}{S\arabic{figure}}
\renewcommand{\figurename}{Supplementary Fig.}

\setcounter{figure}{0}
\renewcommand{\thesection}{S\arabic{section}}
\setcounter{section}{0}

\section*{Supplementary Information}
\section{Details of readout and conversion to state probabilities}
\label{sup:measurement}

\begin{figure}[t]
\includegraphics[width=0.47\textwidth]{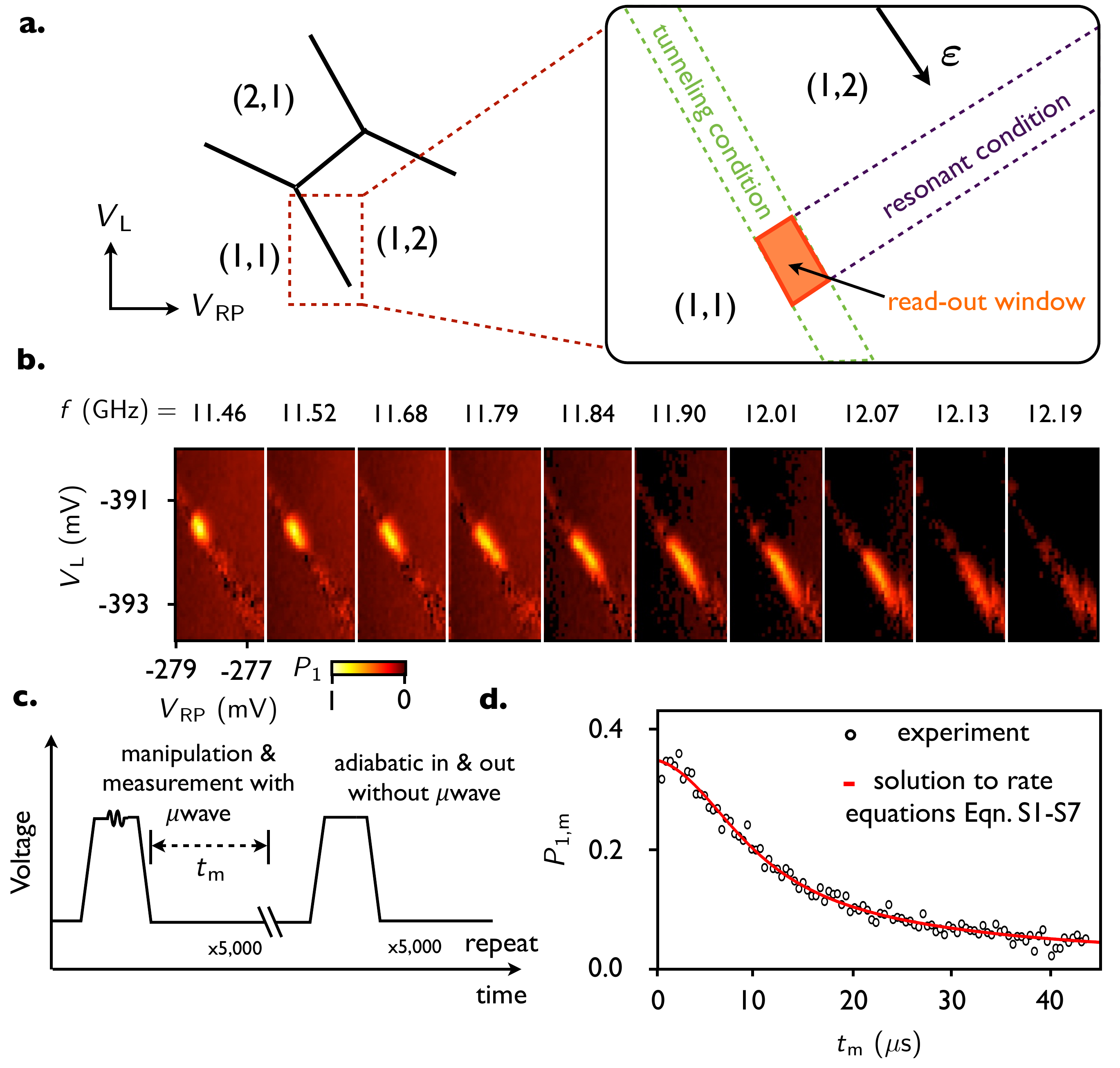}
\caption{ \textbf{Measurement of state populations using energy dependent loading and unloading. a}, Schematic charge stability diagram as a function of gate voltages $V_\text{L}$ and $V_\text{RP}$. The close-up view near the (1,2) to (1,1) charge transition shows a region (orange box) where state manipulation using resonant microwaves as well as robust readout using energy dependent loading and unloading can be performed. In this readout window, applied microwaves are resonant with the qubit energy spacing; in addition, the Fermi level of the right reservoir lies between the qubit states, so that qubit state $\ket{1}$ leads (after tunneling) to a QPC current level corresponding to the (1,1) charge state, whereas state $\ket{0}$ leads to that of (1,2). \textbf{b}, Experimentally measured variation of the readout window as a function of excitation microwave frequency $f$ with a fixed pulse duration of $\approx 5$~ns. The length in gate voltage space of the readout window increases at higher excitation frequency, reflecting the saturation of the qubit level spacing at large detuning. \textbf{c}, Schematic pulse sequence to manipulate the qubit states and measure the population of state $\ket{1}$. 
Note that the adiabatic in- and out- ramps do not induce state population change; this baseline change is included even without microwaves, in order to reduce the background lock-in signal that would otherwise arise from capacitive crosstalk.
\textbf{d}, Measurement of uncalibrated probability of state $\ket{1}$, $P_\text{1,m}$, as a function of the measurement time $t_\text{m}$. The solid red curve shows a fit to the solution of rate equations Eqs.\ S1-S7 with excitation efficiency $\alpha=0.93$, tunnel-out time $T_\text{o}\approx 200$ ns, and reset time $T_\text{i}\approx 2.1\mu s$. }
\label{fig:S1} 
\end{figure}

We use the qubit state-dependent loading and unloading rates to implement robust readout of the state $\ket{1}$ probability. Supplementary Fig.~\ref{fig:S1}a shows a schematic charge stability diagram (left panel) and a magnified region near the (1,1)-(1,2) charge transition (right panel). The green dashed region (tunneling condition) denotes where the Fermi level of the right reservoir lies between the qubit energy levels. In this region, the energy of state $\ket{1}$ is above the Fermi level of the resorvoir, so in this state an electron tunnels out to the right reservoir with characteristic tunneling time $T_\text{o}$, leading to an $I_{QPC}$ level corresponding to the (1,1) charge configuration, whereas state $\ket{0}$ remains in the (1,2) charge state.
In our experiment, the qubit energy splitting is typically $>11$~GHz, at least four times larger than the thermal broadening of the reservoir Fermi level ($\approx$140~mK or $\approx$ 2.8~GHz), which enables high signal contrast between the qubit states. Moreover, it has been shown that the qubit relaxation time $T_{1}$ from $\ket{1}$ to $\ket{0}$ is $>$ 100ms~\cite{Shi:2012p140503}; thus, the readout fidelity and overall coherence time is not limited by $T_{1}$. Excitation from $\ket{0}$ to $\ket{1}$ is performed by applying fixed-frequency microwaves to gate R. The excitation is possible when the detuning is such that the microwaves are resonant with the qubit frequency, which we denote as a purple strip (resonant condition) in Fig.~\ref{fig:S1}a. 
 The intersection of these conditions (the read out window, see the orange box in Fig.~\ref{fig:S1}) defines the region of the stability diagram where the applied microwaves are resonant with the qubit level splitting and also the final state can be read out. Supplementary Fig.~\ref{fig:S1}b shows the variation of the read out window identified by sweeping the voltages on gates L and RP while applying microwave bursts on gate R.

Manipulation sequences, including the microwave bursts, are generated using a Tektronix AWG70002A arbitrary waveform generator (AWG) with a maximum sample rate of 25~Gs/s and an analog bandwidth of about 13~GHz. To generate high frequency ($>$ 10~GHz) microwave signals, we adopted two approaches: (1) For experiments that do not require relative phase control between sequential microwave pulses, we adjust the sample rate of the AWG to generate a sequence of triangular wave segments so that the resulting sequence has a fundamental frequency matching the desired excitation frequency. Higher harmonic components are filtered using a low pass filter with a cutoff frequency of about 18~GHz. Additionally, transmission lines from room temperature electronics to the sample also provide effective attenuation for filtering out $>$ 20~GHz harmonic components. (2) For relative phase control, e.g., the data in Fig. 2d, we increased the effective sampling rate by combining two channels of the AWG, with the second channel output delayed by one half of the sampling time. By constructing odd (even) numbers of sampling points using the first (second) channel of the AWG, we can directly synthesize high frequency $>$ 10~GHz waveform limited only by the analog band width (13~GHz) of the instrument.    

 For measuring changes in qubit state occupation probabilities resulting from fast microwave bursts, we use the general approach described schematically in Fig.~\ref{fig:S1}c, where we measure the difference between the QPC conductance with and without the manipulation pulses. The data are acquired using a lock-in amplifier with a reference signal corresponding to the presence and absence of the pulses (lock-in frequency $\approx$ 777\ Hz). We compare the measured signal level with the corresponding $(1,1)$-$(1,2)$ charge transition signal level, calibrated by sweeping gate L and applying square square pulses with frequency of 777\ Hz to gate L. 
 
To calibrate the state probabilities, we account for the two time scales that affect the overall voltage signal during the measurement phase: (1) the tunneling time $T_\text{o}$ from $\ket{1}$ to (1,1), which determines the onset of the $I_\text{QPC}$ change, and (2) the tunneling time $T_\text{i}$ from (1,1) to $\ket{0}$, which determines the effective reset time. As shown below,  the previously measured~\cite{Shi:2012p140503} $T_{1}$-time of $\approx$~140 ms is much longer then $T_\text{o}$ and $T_\text{i}$; thus, the effect of $T_{1}$ during the measurement phase is negligible. To extract these time scales from experimental measurements at the readout window, as discussed below, we consider the three-state rate equation describing the occupations $n_0(t)$, $n_{1}(t)$, and $n_{(1,1)}(t)$ of the corresponding states $\ket{0}$, $\ket{1}$, and (1,1): 
\begin{equation}
\frac{dn_0(t)}{dt}= n_{(1,1)}(t)/T_i~,
\end{equation}
\begin{equation}
\frac{dn_{(1,1)}(t)}{dt}= -n_{(1,1)}(t)/T_i+n_{1}(t)/T_o~,
\end{equation}
\begin{equation}
\frac{dn_1(t)}{dt}= -n_1(t)/T_o~,
\end{equation}
The boundary conditions appropriate to repeated pulses with duration $t_\text{m}$ are
\begin{equation}
n_0(0)+n_1(0)+n_{(1,1)}(0) = 1~,
\end{equation}
\begin{equation}
n_{(1,1)}(0) = n_{(1,1)}(t_\text{m})~,
\end{equation}
\begin{equation}
n_1(0)=\alpha n_0(t_\text{m})+(1-\alpha) n_1(t_\text{m})~,
\end{equation}
\begin{equation}
n_0(0)=\alpha n_1(t_\text{m})+(1-\alpha) n_0(t_\text{m})~.
\end{equation}
The rate equations describe relaxation of $\ket{1}$ to (1,1) with tunneling time $T_o$ and relaxation of (1,1) to $\ket{0}$ with tunneling time $T_i$; the boundary conditions follow because the microwave pulses convert $\ket{0}$ to $\ket{1}$ with $X_{\pi}$ gate fidelity $\alpha$ and vice versa, and do not affect the occupation of the (1,1) configuration.  

 Experimentally, we measure the uncalibrated probability $P_{1,m}$ of state $\ket{1}$ after an $\text{X}(\pi)$ pulse, which will depend on the measurement time $t_\text{m}$ (see Supplementary Fig.\ \ref{fig:S1}c), and compare the data to the solution of the three-state rate equations. 
Supplementary Fig.~\ref{fig:S1}d shows $P_{1,m}$ as a function of $t_\text{m}$, showing the expected reduction of signal size arising from the reset to state $\ket{0}$. Since the value of $\alpha$ ($X_\pi$ gate fidelity) is not known {\it a priori}, we fit the data to the solution of the rate Eqns. S1-S7 with initial guess of $\alpha$=0.95, and perform the process tomography (QPT) to extract $X_\pi$ gate fidelity. We repeat the fit procedure to self-consistently determine $\alpha$ from the normalization fit and $X_\pi$ gate fidelity from the QPT, with a tolerance in the difference of $\pm$ 0.01, and obtain a converged value of $\alpha$=0.93 after two iterations, consistent with the $X_\pi$ gate fidelity in the main text. The red solid line of Supplementary Fig.~\ref{fig:S1}d shows the converged fit, which yields $T_\text{o}\approx$~200 ns and $T_\text{i}\approx$~2.1 $\mu$s. The probability data in the main text use $t_\text{m} \sim$ 11 $\mu$s, long enough to ensure state reset after the manipulation sequences. Using these time scales allows us to correct for state reset by appropriate normalization of the data.

\section{State tomography of Rabi oscillations in the laboratory frame}
\label{sup:STlab}

\begin{figure}[t]
\includegraphics[width=0.47\textwidth]{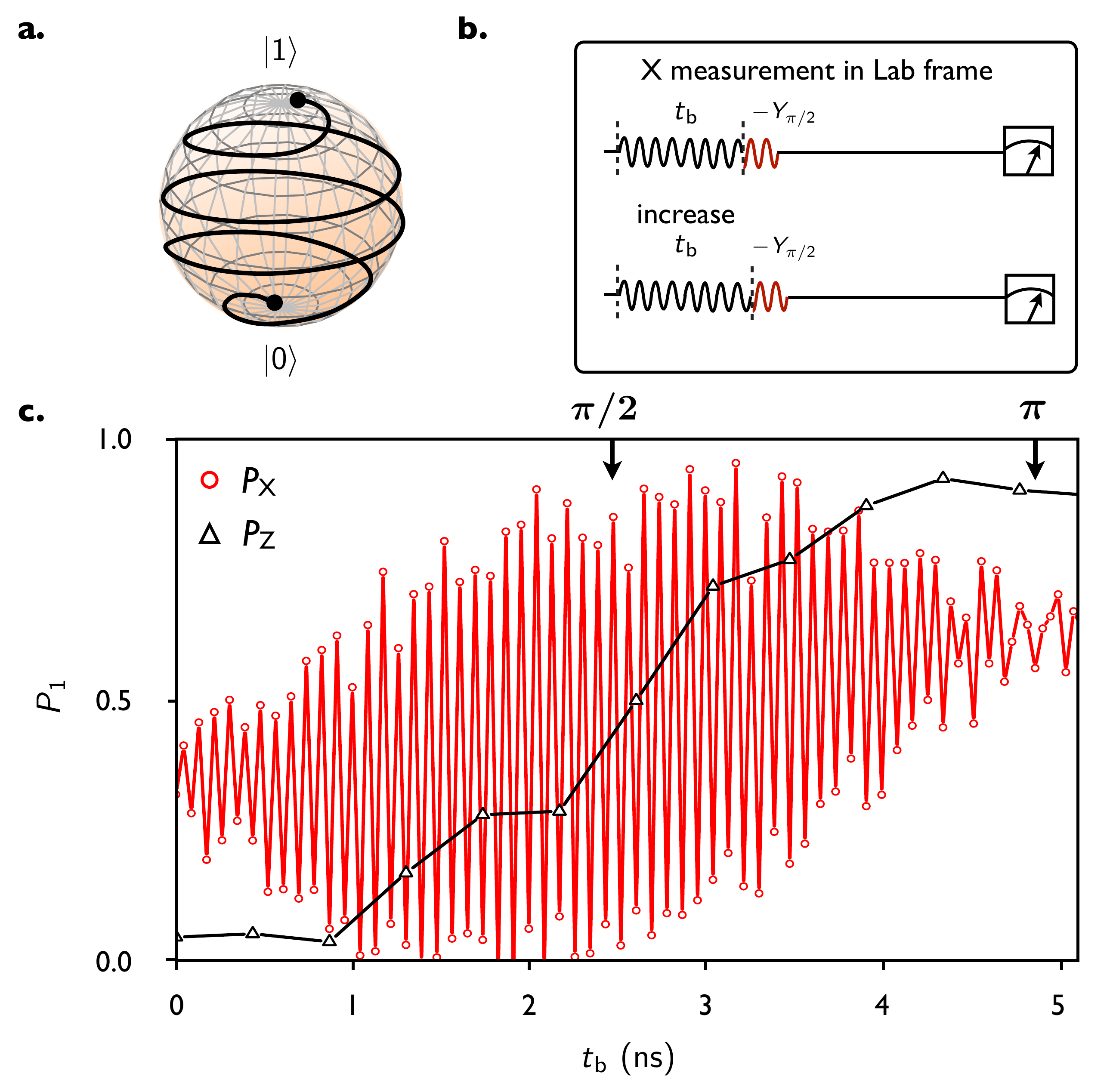}
\caption{\textbf{State tomography of the Rabi oscillation in the laboratory frame. a}, Schematic Bloch sphere representation of the expected state evolution upon resonant microwave drive in the laboratory frame. \textbf{b}, Microwave pulse sequences used to measure the X-axis projection of the Bloch vector during the Rabi oscillations. Note that the measurement pulse $(-Y_{\pi/2})$ has a fixed phase shift of -90 degrees with respect to the start of the Rabi drive pulse, which has length $t_\text{b}$. \textbf{c}, X-axis (red circles) and Z-axis (black triangles) projection of the Bloch vector during a Rabi pulse obtained by lab frame tomography measurements as a function of $t_\text{b}$. The data is shown as a function of $t_\text{b}$ up to a time corresponding to an $X_\pi$ rotation. The X-axis projection is clearly consistent with a spiral trajectory of the Rabi nutation in the lab frame, which is shown schematically in $\textbf{a}$.}
\label{fig:S2} 
\end{figure}

The state tomography of Rabi oscillations in the main text are performed in the rotating frame. Here we also show tomographic measurements in the lab frame. Supplementary Fig.~\ref{fig:S2}a shows a schematic of the evolution of the Bloch vector upon resonant microwave excitation. To implement measurement in the lab frame after microwave manipulation, we fix the phase of the measurement $\pi/2$ pulse and apply it immediately after a Rabi drive pulse of length $t_\text{b}$. Supplementary Fig.~\ref{fig:S2}b shows, as an example, the microwave sequence for the X-axis projection measurement of the continuous Rabi oscillation. Note that Larmor phase accumulation does not affect the Z-axis projection measurement, so that the rotating and lab frame measurements of the Z-axis component of the Bloch vector are identical. The result of the lab frame state tomography is shown in Supplementary Fig.~\ref{fig:S2}c. The rapid oscillation of the X-axis projection $P_\text{x}$ of the probability varies at the qubit frequency of $\approx 11.52$ GHz, starts and ends at probability near 1/2, and is consistent with the expected time evolution of the qubit state under resonant microwave drive, showing good control over the measurement axis in this experiment. Moreover, measurement of state evolution in the lab frame can be a useful tool to understand decoherence of a driven qubit~\cite{Jing:2014p022118, Yan:2013p2337}. The resolution (i.e., the point density) of the measurement of this rapid precession is currently limited by the sampling time of the arbitrary waveform generator (Tektronix 70002A, set to 43.4 ps).

\bibliography{siliconqcsncwin}

\end{document}